# Sintesi di algoritmi con SKY

Giovambattista Ianni
Dipartimento di Matematica
Università della Calabria
87036, Rende (CS), Italia
ianni@deis.unical.it

## 1 Introduzione

In questo lavoro viene proposta una estensione del linguaggio DATALOG$^{CIRC}$[CP97]. DATALOG$^{CIRC}$ è un linguaggio logico basato su DATALOG, la cui innovazione sostanziale consiste nell'aumento della potenza espressiva, tramite l'introduzione di una semantica basata sulla circoscrizione, un meccanismo deduttivo non monotòno che generalizza l'ipotesi di mondo chiuso.

Questo tipo di semantica fornisce interessanti vantaggi teorici, ma al tempo stesso non consente che la valutazione di un programma DATALOG$^{CIRC}$ sia affrontabile dal punto di vista pratico, a causa dell'enorme numero di modelli minimali che un possibile algoritmo di risoluzione dovrebbe prendere in considerazione. Nonostante questo, DATALOG$^{CIRC}$ mette in grande risalto alcune caratteristiche che potrebbero essere utilizzate come modello per un linguaggio di specifica per la sintesi di algoritmi basati su tecniche algoritmiche generali.

Prenderemo soprattutto in considerazione il backtracking, una tecnica algoritmica principalmente fondata sul principio di raffinamento di una soluzione parziale, e su criteri di sfrondatura dell'albero di ricerca che si viene a determinare. Si tratta della naturale evoluzione delle semplici tecniche di enumerazione dello spazio delle soluzioni.

In questo lavoro presenteremo quindi SKY : un linguaggio che estendendo DATALOG$^{CIRC}$, permette un utilizzo più intelligente e trasparente delle potenzialità di quest'ultimo, e nel contempo si propone come potente strumento per la specifica e la generazione di algoritmi risolutivi a struttura sia enumerativa che tramite backtracking.

## 2 Sintassi e semantica di SKY

In questa sezione definiremo in modo informale ma, laddove necessario, rigoroso, la sintassi e la semantica di un programma SKY. Nella sottosezione 2.1 provvederemo a definire la sintassi del linguaggio e alcuni suoi aspetti semantici non centrali; passeremo poi ad affrontare i dettagli più delicati della semanti-ca di SKY nelle sottosezioni 3.2 –dove ci occuperemo della struttura portante del motore di valutazione Semi-naïf$^+$– e 3.3, dove mostreremo dettagliatamente cosa avviene in ogni singola iterazione di Semi-naïf$^+$.

## 2.1 Sintassi e primi aspetti semantici

Un programma SKY è costituito da 4 sezioni principali:

- La sezione **BOUNDS**, dove si può, e in alcuni casi si deve, specificare l'universo di valutazione delle singole relazioni.

- La sezione **TEMPLATE** che raccoglie spezzoni di programma generici adatti al riutilizzo.

- La sezione **GENERATE**, il corpo vero e proprio del programma, che si occupa di specificare e/o generare il dominio delle soluzioni parziali ammissibili.

- La sezione **CHECK**, che rappresenta il programma di verifica di ammissibilità delle soluzioni parziali.

### 2.1.1 La sezione GENERATE

La sezione **GENERATE** è individuata da una intestazione del tipo

$$MAIN < pred_1(\_,...,\_),...,pred_n(\_,...,\_) >$$

dove la serie di predicati $pred_1 ... pred_n$ indica il nome delle relazioni di input; un eventuale sequenza di underscore, '$\_$', separati da virgole e contenuta tra parentesi tonde subito dopo il nome del predicato, ne indica l'arità. E' possibile anche la sintassi $pred_k(number)$ dove il valore numerico $number$ specifica direttamente l'arità attesa per $pred_k$. Segue un elenco di clausole che da questo momento chiameremo *clausole estese*. Una clausola estesa possiede formalmente la struttura di una regola di Horn classica:

$$head(\mathbf{X}) \leftarrow costruttore_1,...,costruttore_n$$

Dove con la notazione $\mathbf{X}$, se non specificato diversamente, indichiamo un elenco di cosiddetti *argomenti estesi*, separati da virgole, che prenderemo in esame nel seguito. I predicati presenti nella parte antecedente di una regola estesa (che chiameremo d'ora in poi *predicati estesi* o *costruttori*), possono essere:

- Un predicato di Horn classico, che è inoltre l'unico tipo consentito per i predicati di testa;

- Un costruttore di intervallo.

- Un costruttore di iterazione.

- Un costruttore di template.

- Un costruttore di complementazione semplice.
- Una espressione booleana.

La premessa di una regola può essere eventualmente assente, e in questo caso una regola assume la forma di *fatto*:

$$head(\mathbf{X}).$$

### 2.1.2 Costruttori di iterazione

Un costruttore di iterazione è uno tra i seguenti:

$$
\begin{array}{rll}
\textbf{Subset} & (\mathbf{Y}) & [predicato(\mathbf{X})] \\
\textbf{Range} & (\mathbf{Y}) & [predicato(\mathbf{X})] \\
\textbf{Something} & & (\mathbf{Y})(\mathbf{X}) \\
\textbf{Any} & (\mathbf{Y}) & [predicato(\mathbf{X})] \\
\textbf{Permutation} & (\mathbf{Y}) & [predicato(\mathbf{X})](Tag) \\
\textbf{Partition} & (\mathbf{Y}) & [predicato(\mathbf{X}), Cardinality](Tag) \\
\textbf{Co}^* & & [predicato(\mathbf{X})]
\end{array}
$$

Dove $Tag$ è un singolo argomento esteso. Chiamiamo *argomenti di split* quelli appartenenti a $\mathbf{Y}$, e *argomenti di iterazione* quelli appartenenti a $\mathbf{X}$, mentre chiameremo *predicato di origine* il predicato compreso tra parentesi quadre per ciascun costruttore (che deve appartenere all'insieme dei predicati di input). Si noti che $Co*$ è solo sintatticamente simile a un costruttore di iterazione, ma il suo trattamento semantico è abbastanza diverso.

### 2.1.3 Costruttori di template

Un template viene costruito tramite una dichiarazione del tipo:

$$templatename < predicato_1(\mathbf{X}_1), \ldots, predicato_n(\mathbf{X}_n) > (\textbf{Args}) \qquad (1)$$

Il motore di valutazione deve essere in grado di risalire –secondo la strategia che vedremo nella sezione 3.1.3– a una corrispondente definizione nella sezione **TEMPLATE**, esclusi i template di libreria immediatamente disponibili. Ogni argomento presente in una delle sequenze di argomenti $\mathbf{X}_i$ può essere sostituito da un underscore '_', o da un asterisco '*'.

#### 2.1.3.1 Costruttore di complementazione semplice
Il metapredicato $Co$ è sintatticamente identico al suo omologo $Co^*$:

$$\textbf{Co}[predicato\_esteso(\mathbf{X})]$$

Come vedremo nel seguito le regole che fanno uso di $Co$ devono essere stratificate rispetto a questo metapredicato.

### 2.1.4 Costruttore di intervallo

Un costruttore di intervallo viene invocato nel modo seguente:

$$\{lowerarg..upperarg\}(arg)$$

dove *lowerarg* e *upperarg* possono essere delle costanti numeriche o un indicazione di conteggio di cardinalità del tipo $count <p>$, che fissa l'estremo relativo alla cardinalità di $p$; $p$ deve essere un predicato di input.

### 2.1.5 Argomenti estesi

Un argomento esteso può essere:

- un nome di variabile, identificato dalla lettera iniziale maiuscola.
- un nome di costante, identificato dalla lettera iniziale minuscola
- una espressione numerica intera, comprendente nomi di variabili, costanti numeriche intere e gli operatori $*$, $/$, $+$, e $-$.

### 2.1.6 Espressioni booleane

Una espressione booleana è costituita da due espressioni aritmetiche intere separate dagli operatori $<$, $>$, $\leq$, $\geq$, $=$ e $\neq$,

## 2.2 La sezione CHECK

Questo sottoprogramma è costituito da un insieme di regole DATALOG standard, non ricorsive. In particolare possono essere presenti qui delle regole contenenti in testa i letterali speciali $fail$ e $fail^*$. E' consentito l'uso del metapredicato di complementazione $CO$.

## 2.3 La sezione BOUNDS

Nella sezione **BOUNDS** è possibile indicare il dominio di valutazione di ciascun predicato. E' composta da un elenco di regole DATALOG standard, con l'aggiunta dei costruttori di intervallo. Nel corpo possono essere usati esclusivamente relazioni EDB; un costruttore di intervallo può avere come estremo una costante e/o il template '$count <p>$' dove $p$ deve essere un predicato di input.

## 2.4 La sezione TEMPLATES

Questa sezione è costituita da un'elenco di *dichiarazioni di template* costituite nel seguente modo:

- Un'intestazione di template, dove sono elencati i parametri formali di ingresso (nomi di predicati con indicazione di arità), e l'arità di uscita del predicato risultante:

  **template** $templatename < pred_1(arity), ..., pred_n(arity) > (arity)$

  dove *arity* può essere sia una costante numerica intera o un elenco di underscore '_' separati da virgole.

- un elenco di regole estese: deve esistere obbligatoriamente almeno una regola con in testa *templatename*; è consentito invocare un costruttore di template, ma è vietato fare uso di ricorsione tra chiamate di template sia dirette che indirette. Non è consentita la presenza di nessuno tra i $pred_i$ in testa a una qualche regola.

# 3 L'algoritmo Semi-naïf$^+$

L'algoritmo Semi-naïf$^+$è una variante del noto metodo di valutazione differenziale per query in DATALOG[Ull91] [BR87]. Ne daremo qui per esteso una definizione formale. Quest'algoritmo ricalca sostanzialmente le linee di un generico sistema di risoluzione in backtracking, dove gli spazi di ricerca sono specificabili dall'utente tramite l'uso degli iteratori.

## 3.1 Da SKY a SKY$^{Plain}$

In questa sezione mostreremo il comportamento di molti dei costrutti di contorno di SKY (template,bounds,aritmetica,negazioni $CO^*$) generando opportuni insiemi di regole equivalenti, e modificando eventualmente le regole già esistenti: il risultato finale che ci proponiamo di ottenere è di passare da un programma SKY a un programma espresso in SKY$^{Plain}$. Un programma SKY$^{Plain}$ è essenzialmente un programma SKY dove sono presenti le sole sezioni *generate* e *check*, e ciascuna regola contiene solo atomi canonici e/o costruttori di iterazione. Le definizioni formali che daremo qui sono del tutto generali e per questo ancora inefficienti, ma costituiscono le necessarie linee guida che una possibile implementazione deve rispettare per avere una semantica equivalente. Dove non indicato esplicitamente, le nuove regole generate nel trasformare un programma SKY in SKY$^{Plain}$ si considerano aggiunte alla sezione *generate*.

### 3.1.1 Esplosione delle espressioni aritmetiche

Daremo qui le indicazioni a cui deve attenersi un implementazione per disporre di aritmetica su numeri interi positivi, a partire dallo 0. La semantica che mostreremo (sicuramente la meno efficiente) mostra come si possa impedire l'insorgere di problemi teorici piuttosto gravi, quali la non terminazione in tempo finito della valutazione di un programma SKY. Anche in questo caso suggeriamo che nella implementazione si realizzi un sistema di calcolo più leggero con reale possibilità di utilizzo pratico.

Supponiamo dunque di trovare un argomento esteso che sia un'espressione aritmetica, sia questo *expr*. Anzitutto dobbiamo stabilire quale sia il sistema aritmetico a cui questa espressione fa riferimento:

- Se l'espressione aritmetica si trova all'interno di un predicato *pred* in posizione $i$, questo deve possedere obbligatoriamente una regola BOUND $b$. Questa regola è immediatamente valutabile tramite algebra relazionale, essendo composto il corpo esclusivamente da predicati intensionali. Sia $B$ la relazione risultato della valutazione. Allora il sistema aritmetico di *expr* sarà limitato all'intervallo

$$0, \ldots, MAX(\prod_i B)$$

  Dove si è fatto l'assunto che tutte le costanti di $\prod_1 B$ siano valori interi numerici positivi.

- In tutte le altre situazioni (espressioni booleane, argomenti **Y** e $\alpha$ di costruttori di iterazione, argomenti di uscita di template) sia **X** l'insieme (che deve essere non vuoto) costituito da:
    1. costanti numeriche di *expr*.
    2. variabili in *expr* che compaiono anche in predicati ordinari nella stessa regola, che posseggano una regola BOUND.

  sia allora $\overline{\mathbf{X}}$ l'insieme degli 'upper bound' relativi ai sistemi numerici di ciascun elemento di **X**, determinati per ciascun elemento $x_i$ di **X** come segue:

    1. Se $x_i$ è un costante numerica allora il suo sistema aritmetico è limitato da

$$0, \ldots, x_i$$

    2. Se $x_i$ è una variabile allora il suo sistema aritmetico è limitato a

$$0, \ldots, MAX(MAX(\prod_\alpha B_1), \ldots, MAX(\prod_\gamma B_n))$$

    dove $B_1, \ldots, B_n$ sono le relazioni risultato della valutazione delle regole bound associate ai predicati dove compare $x_i$ nella stessa regola in esame.

  Allora il sistema aritmetico complessivo sarà limitato a:

$$0, \ldots, MAX(\overline{\mathbf{X}})$$

A questo punto siamo in grado di stabilire tutti i possibili sistemi aritmetici utilizzati nel programma. Per ciascun sistema aritmetico $\Phi$ (limitato da 0 a $\phi$) dovremo aggiungere tutte le seguenti regole e fatti:

$next_\Phi(0,1)$.
$next_\Phi(1,2)$.
...
$next_\Phi(\phi-1,\phi)$.
$sum_\Phi(X,0,X)$.
$sum_\Phi(X,Y,K) \leftarrow sum_\Phi(Y,X,K)$
$sum_\Phi(X,1,K) \leftarrow next_\Phi(X,K)$
$sum_\Phi(X,Y,K) \leftarrow sum_\Phi(Y_1,1,Y), sum_\Phi(X,Y_1,K_1), sum_\Phi(K_1,1,K)$
$prod_\Phi(X,1,X)$.
$prod_\Phi(X,0,0)$.
$prod_\Phi(X,Y,K) \leftarrow prod_\Phi(Y,X,K)$
$prod_\Phi(X,Y,K) \leftarrow prod_\Phi(X,Y_1,K_1), sum_\Phi(X_1,1,X), sum_\Phi(K_1,Y,K)$

Si noti che utilizzando $prod_\Phi$ come relazione inversa non esiste per il momento arrotondamento. Fatto questo, dovremo modificare le regole contenenti espressioni in maniera tale da utilizzare i predicati appena inseriti nel programma. Supponiamo quindi di avere una espressione $e$ e di avere stabilito che il suo sistema aritmetico è $\Phi$. Sia $r$ la regola del programma in cui compare questa espressione. L'esplosione delle espressioni aritmetiche è articolata nei seguenti punti:

1. Una espressione può essere considerata essenzialmente un termine più una sommatoria di *termini* (eventualmente vuota): cioè sia

$$e \equiv \Gamma_1 \pm_k \sum_{i=2}^{n} \Gamma_i$$

aggiungeremo quindi come primo predicato della parte destra di $r$,

- $sum_\Phi(\Gamma_1, \sum_{i=2}^{n} \Gamma_i, \Lambda)$ se $\pm_k$ è un segno $+$;
- $sum_\Phi(\Lambda, \sum_{i=2}^{n} \Gamma_i, \Gamma_1)$ se $\pm_k$ è un segno $-$;

mentre dove occorre $e$ si rimpiazzi la variabile $\Lambda$. Si ripeta ricorsivamente questo punto su $\sum_{i=1}^{n} \Gamma_i$, che è evidentemente ancora una espressione, fino a che i sottoobiettivi $sum_\Phi$ non presentano come argomenti esclusivamente termini;

2. Evidentemente è ora necessario sapere come ridurre un termine: un termine $t$ può essere considerato come un fattore moltiplicato (o diviso) per una produttoria di *fattori* (eventualmente vuota): cioè sia

$$t \equiv \Xi_1 \diamond_k \prod_{i=2}^{n} \Xi_i$$

aggiungeremo quindi come primo predicato della parte destra di $r$,

- $prod_\Phi(\Xi_1, \prod_{i=2}^n \Xi_i, \Theta)$ se $\diamond_k$ è un segno $\times$;
- $prod_\Phi(\Theta, \prod_{i=2}^n \Xi_i, \Xi_1)$ se $\diamond_k$ è un segno $/$;

mentre dove occorre $t$ si rimpiazzi la variabile $\Theta$. Si ripeta ricorsivamente questo punto su $\prod_{i=1}^n \Xi_i$, che è ancora un termine, fino a che i sottoobiettivi $prod_\Phi$ non presentano come argomenti esclusivamente fattori;

3. Se occorre ridurre un fattore $f$ allora basta tenere conto che questo può essere:

   - una costante o una variabile, per cui non è necessario compiere alcuna operazione;
   - una espressione delimitata dalle parentesi '(' e ')', e in questo caso si ripetono ricorsivamente tutte le operazioni necessarie a partire dal punto 1.

Per esempio, supponiamo di dovere esplodere la seguente regola:

$$avg((X_1 + X_2)/2) \leftarrow n(X_1), n(X_2)$$

In sequenza verrebbero operati i seguenti passaggi:

$$\begin{aligned} avg(X_1, X_2, (X_1 + X_2)/2) &\leftarrow n(X_1), n(X_2) \Longrightarrow \\ avg(X_1, X_2, Y) &\leftarrow prod(2, Y, X_1 + X_2), n(X_1), n(X_2) \Longrightarrow \\ avg(X_1, X_2, Y) &\leftarrow prod(2, Y, Z), sum(X_1, X_2, Z), n(X_1), n(X_2) \end{aligned}$$

### 3.1.2 Esplosione dei predicati CO*

Nel caso in cui nel programma SKY si sia costretti a utilizzare il metapredicato di complementazione generale, SKY sarà costretto a calcolare la relazione complemento per via iterativa. Nel programma $\mathsf{SKY}^{Plain}$ ciò comporta l'aggiunta di un certo numero di nuove regole.

Per ogni predicato pred, se nel programma $\mathsf{SKY}^{Plain}$ compaiono una o più occorrenze di un metapredicato $CO^*[pred(\mathbf{X})]$ allora sostituiremo ciascuna occorrenza con il predicato $CO\_pred(\mathbf{X})$ e aggiungeremo nella sezione *check*, le regole:

$$\begin{aligned} \overline{\overline{CO\_pred}}(\mathbf{X}) &\leftarrow pred(\mathbf{X}) \\ \overline{\overline{CO\_pred}}(\mathbf{X}) &\leftarrow CO\_pred(\mathbf{X}) \\ fail^* &\leftarrow pred(\mathbf{X}), CO\_pred(\mathbf{X}) \end{aligned}$$

e nella sezione *generate* la regola:

$$\overline{CO\_pred}(\mathbf{X}) \leftarrow \mathbf{Something}(\mathbf{X})$$

### 3.1.3 'Inflazione' dei template

Questa sezione dell'algoritmo Semi-naïf$^+$ ha il compito di esplodere i descrittori di template eventualmente usati, aggiungendo al programma le regole che effettivamente specificano il sottoprogramma che nasce dall'unione tra definizione di template e una occorrenza di costruttore di template.

Ogni template prevede al suo interno tutte le sezioni che può avere programma principale. Supporremo che ogni volta che si prende in considerazione una regola appartenente a una certa sezione, il gruppo di regole che si ricaverà andrà posto nella stessa sezione del programma principale. Se non si specifica nessuna sezione in particolare, le regole esplose andranno incluse nella sezione *generate* del programma principale.

Ricevuto il programma SKY $\mathcal{P}$ elaborato dalle fasi di preparazione precedenti, per ogni costruttore di template $t$ occorrente in qualche regola $r_p$ di $\mathcal{P}$ nella forma

$$\mathbf{t} < actual_1(\mathbf{X}_1^{act}), ..., actual_n(\mathbf{X}_n^{act}) > (\mathbf{Args}) \qquad (2)$$

Se in qualche $\mathbf{X}_i^{act}$ dovessero comparire degli argomenti asteriscati, aggiungeremo al programma $\mathcal{P}$ la regola

$$actual'_i(\mathbf{X}_i'^{act}) \leftarrow actual(\mathbf{X}_i^{act})$$

dove $\mathbf{X}_i'^{act}$ è una sequenza di argomenti privata dei campi asteriscati.

In un caso del genere, senza ledere la generalità, supporremo di aver aggiunto una tale regola e ci comporteremo come se nella 2 non ci siano argomenti asteriscati, immaginando di aver sostituito, ove necessario $actual_i(\mathbf{X}_1^{act})$ con $actual'_i(\mathbf{X}_1'^{act})$. Chiamiamo $\mathbf{X}_i^F$ l'insieme degli indici posizionali delle variabili indicate con '_' in ciascun $\mathbf{X}_i^{act}$ (e individuati singolarmente come $\mathbf{X}_i^{Fj}$), e $\mathbf{X}_i^B$ l'insieme, disgiunto dal primo, degli indici rimanenti (i cui elementi singoli saranno individuati da $\mathbf{X}_i^{Bj}$); diremo $\mathbf{X}_{tot}^F$ la sequenza complessiva di tutte le coppie $< actual_i, \mathbf{X}_i^{Fj} >$ per ogni $i$ che va da 1 a $n$ e, a parità di $i$, per ogni $j$ che va da 1 a $|\mathbf{X}_i^F|$. Costruiremo e aggiungeremo al programma $\mathcal{P}$ l'insieme di regole $\mathcal{R}_t$ associato a ciascun costruttore di template presente in $\mathcal{P}$, generato secondo il seguente criterio:

supponiamo di avere a disposizione la definizione formale associata a un costruttore di template 2, e sia questa

$$\mathbf{template\ t} < formal_1(arity_1), ..., formal_n(arity_n) > (arity) \qquad (3)$$

assumeremo che ciascun $arity_i$ sia pari a $|\mathbf{X}_i^F|$. Inoltre, diremo $\mathcal{T}$ l'insieme di regole che specificano nel programma $\mathcal{P}$ la 3.

Per ogni regola $r_t$ di $\mathcal{T}$, se questa contiene costruttori di template, vengano esplosi ricorsivamente secondo la stessa procedura che andiamo a descrivere.

Data la regola $r_t^{PLAIN}$, priva di costruttori di template, risultante aggiungeremo a $\mathcal{R}_t$ una corrispondente regola $r_t'$ costruita partendo da $r_t^{PLAIN}$ secondo le seguenti indicazioni:

1. Qualunque occorrenza $pred_i(\mathbf{X}_i)$ in $r_t^{PLAIN}$ che faccia riferimento a un qualsiasi predicato utilizzato al di fuori della definizione del template in causa viene lasciata inalterata in $r_t'$ [1], cioè il predicato in questione viene trattato come predicato globale.

2. Ad ogni occorrenza di un qualche $formal_i(\mathbf{X}_i)$ in $r_t^{PLAIN}$, sostituiremo il predicato $actual_i(\mathbf{X}_i')$. La sequenza di argomenti $\mathbf{X}_i'$ viene così composta: per ogni j che va da 1 a $|\mathbf{X}_i^B|$ metteremo in posizione $\mathbf{X}_i^{Bj}$ l'argomento nella medesima posizione prelevato da $\mathbf{X}_i^{act}$. Per ogni j che va da 1 a $|\mathbf{X}_i^F|$ metteremo in posizione $\mathbf{X}_i^{Fj}$ di $\mathbf{X}_i'$ l'argomento j-esimo di $\mathbf{X}_i$.

3. Ogni altra occorrenza di predicato $pred_i(\mathbf{X}_i)$ in $r_t^{PLAIN}$, incluse le occorrenze di $\mathbf{t}$ e indifferentemente se ci si trova in testa o in coda, viene sostituita con $pred_i'(\mathbf{X}_i')$. Il nome $pred_i'$ è unico per ogni occorrenza di $pred_i$ in $r_t$ e in tutte le regole di $\mathcal{T}$, ma diverso per ogni diversa occorrenza del tipo 2 nel programma $\mathcal{P}$ che richieda di essere esplosa come qui viene descritto. $\mathbf{X}_1'$ sarà identico, argomento per argomento, a $\mathbf{X}_i$ per le prime $|\mathbf{X}_i|$ posizioni, mentre per ogni j che va da 1 a $|\mathbf{X}_{tot}^F|$ metteremo nella posizione $|\mathbf{X}_i| + j$ di $\mathbf{X}_i'$ gli argomenti associati alla coppia $\mathbf{X}_{tot}^{Fj}$ di $\mathbf{X}_{tot}^F$.

4. Si modifichi infine la regola $r_p$ contenente l'occorrenza di costruttore 2, sostituendo a questa $\mathbf{t}'(\mathbf{X}_t')$ dove $\mathbf{X}_t'$ è ottenuta come al punto precedente si ricava $\mathbf{X}_i'$, basandosi però su **Args** anzichè su $\mathbf{X}_i$. Anche in questo caso il nome $\mathbf{t}'$ è da considerarsi diverso per ciascuna occorrenza 2 sottoposta alla procedura di esplosione.

Riassumendo, questa fase esplode ed elimina ogni occorrenza di costruttore di template in $\mathcal{P}$ aggiungendo a questo un certo insieme di regole $\mathcal{R}_t$ e modificando la regola in cui si verifica l'occorrenza data, in base alla definizione formale di template corrispondente. Il processo viene ripetuto ricorsivamente su $\mathcal{P}$ finchè questo non è completamente privo di costruttori di template. Come già detto, per garantire la terminazione di questa fase, non è consentito l'uso di ricorsione diretta o indiretta tra template.

**Esempio 3.1** *Supponiamo di avere a disposizione nella sezione* **Template** *di un nostro programma la seguente definizione:*

---

[1]Sono escluse da questo punto le relazioni aritmetiche e di libreria.

$$\textbf{template } JoinAndProject < p1(2), p2(1) > (\_)$$

$$JoinAndProject(X) \leftarrow \_p1(X,Y), \_p2(Y)$$

Potremo invocare questo template con una regola

$$fail \leftarrow JoinAndProject < edge(\_,\_), node(\_) > (pf)$$

L'operazione di inflazione su questo costruttore sostituirà questa regola con:

$$fail \leftarrow JoinAndProject0000(pf)$$

mentre il programma complessivo si arricchirà della regola

$$JoinAndProject0000(X) \leftarrow edge(X,Y), node(Y)$$

Meno banale può essere uno uso di costruttore come in questa regola

$$fail \leftarrow JoinAndProject < edge(\_,\_), edge(ff,\_) > (kk)$$

La nuova regola che rimpiazza la precedente è

$$fail \leftarrow JoinAndProject0001(kk, ff)$$

mentre viene aggiunta la regola

$$JoinAndProject0001(ff, X) \leftarrow edge(X,Y), edge(ff,Y)$$

Come si vede l'uso dei template è spesso non banale, e risulta sicuramente gratificante per un utente esperto che abbia necessità di sintetizzare al meglio una specifica SKY. Si noti che per il momento si è volutamente trascurata la questione riguardante un possibile uso di variabili locali nella definizione di template. E' ugualmente importante fare notare che lo stesso vale per i nomi di predicato. Siccome l'operazione di inflazione è diversa a seconda della globalità o meno di un predicato è buona norma far precedere i predicati locali da un '_'. Una invocazione di costruttore del tipo

$$fail \leftarrow JoinAndProject < edge(\_,\_), edge(X,\_) > (kk)$$

da luogo a

$$JoinAndProject0001(X, X) \leftarrow edge(X,Y), edge(X,Y)$$

raccomandiamo di utilizzare questa possibilità con cautela, poichè non sempre potrebbe dare luogo all'effetto desiderato.

### 3.1.4 Esplosione della sezione BOUND

Una comodità che SKY offre è quella di limitare automaticamente ogni predicato, se richiesto. Il codice di specifica diventa così più sintetico: molto spesso non c'è bisogno di garantire esplicitamente la *sicurezza* di una regola se si usa la negazione, e inoltre una implementazione intelligente può usare queste informazioni per ridurre il numero di operazioni da svolgere nella valutazione di ciascuna regola. Noi ci limitiamo qui a spiegare semanticamente che cosa comporta per ciascun predicato dichiarare un bound.

Consideriamo ora ciascuna regola presente nella sezione Bound. Chiamiamo $b$ una di queste regole, e supponiamo siano della forma

$$pred(\mathbf{X}) \leftarrow pred_1(\mathbf{X}_1), \ldots, pred_n(\mathbf{X}_n)$$

dove i predicati $pred_1, \ldots, pred_n$ sono da considerare esclusivamente predicati di input, ma senza costruttori di iterazione, e senza alcun tipo di ricorsione. Aggiungiamo al programma $\mathsf{SKY}^{Plain}$ la regola

$$bpred(\mathbf{X}) \leftarrow pred_1(\mathbf{X}_1), \ldots, pred_n(\mathbf{X}_n)$$

Chiamiamo *bpred predicato bound* di *pred*. Per esplodere completamente la regola $b$ cerchiamo ogni regola della sezione **MAIN** che ha in testa *pred*: chiamiamo questa regola $c$ e sia questa

$$pred(\mathbf{X}') \leftarrow pred'_1(\ldots), \ldots, pred'_m(\ldots)$$

e la modifichiamo nel seguente modo:

$$pred(\mathbf{X}') \leftarrow bpred(\mathbf{X}'), pred'_1(\ldots), \ldots, pred'_m(\ldots)$$

Inoltre cerchiamo tutte le regole dove è presente nel corpo la complementazione $CO$, ma che non sono safe, del tipo

$$p(\mathbf{X}) \leftarrow p_1, \ldots, p_k, CO[p_{k+1}(\mathbf{X})], CO[p_m(\mathbf{X}_m]$$

per ogni atomo $p_k(\mathbf{X})$ che è presente come negato nel corpo, se in $\mathbf{X}$ è presente almeno una variabile non limitata, allora si consideri il predicato bound $bp_k$ relativo a $p_k$ e si inserisca nel corpo l'atomo $bp_k(\mathbf{X})$.

## 3.2 Struttura globale di Semi-naïf$^+$

Inizialmente si è pensato ad un algoritmo di valutazione enumerativo. Questa modalità di valutazione è sostanzialmente un algoritmo enumerativo su tutte le possibili istanze di predicati $Q$ generabili sull'universo delle costanti in gioco. Un primo passo avanti è stato fatto da [Vas98], dove con il linguaggio FULL-SPEC ci si è prefissi sostanzialmente l'obiettivo di sottoporre la struttura enumerativa a un carico più leggero, realizzato introducendo dei metapredicati capaci di specificare con più precisione le istanze di predicati $Q$ effettivamente da prendere in considerazione. SKY estende questa possibilità: l'utente può non solo specificare un dominio di valutazione adeguato per il problema per cui intende sintetizzare un algoritmo, ma, se vuole, può anche specificare in vari modi come devono essere costruite le soluzioni potenzialmente ammissibili da valutare.

Un programma SKY, a differenza di DATALOG$^{CIRC}$ non possiede direttamente una query da verificare. E' possibile mostrare che il sistema di valutazione complessivo si comporta tuttavia come se dovesse soddisfare una particolare query ristretta[2], non direttamente visibile all'utente del linguaggio. Segue ora la descrizione dell'algoritmo Semi-naïf$^+$:

**Algoritmo 3.1** *(Semi-naïf$^+$.)* L'algoritmo Semi-naïf$^+$riceve in input un programma SKY che assumeremo corretto e ritorna in output 'NO' o 'YES'. In quest'ultimo caso si può restituire un'insieme di relazioni che certifica il perchè della risposta positiva. Supporremo inoltre che $E^2$ sia stato inizializzato in base al programma SKY dato

PROGRAM Semi-naïf$^+$;

**Valuta la sezione *generate* una prima volta**;
**Do**
  **Valuta la sezione *check***;
  do_backtracking := EE.**fail**;
  maybe_found := ¬ EE.**fail\*** ∧ EE.ReachedFixedPoint ∧ $(\forall i \forall \mathbf{X}(\overline{\overline{CO}}_i(\mathbf{X}))$;
  **If** (maybe_found ∧¬ do_backtracking)
    **OutputYES**;
    **Exit**;
  **If** (do_backtracking)
    **Do**
      do_backtracking = **NOT** EE.ActualIterator++;
    **While** (do_backtracking **AND NOT** EE.Iterators.empty())
    **Valuta ancora la sezione *generate***;
**While NOT** EE.Iterators.empty();
**OutputNO**;

Come si vede la struttura base di Semi-naïf$^+$ consiste in un particolare ciclo che principalmente non fa altro che pilotare $E^2$(*Evaluation Engine* motore di valutazione), il cuore dell'algoritmo. Bisogna subito rimarcare che le modalità di valutazione della sezione *check* sono diverse dalle modalità di valutazione della sezione *generate*. La prima è costituita da un insieme di regole non ricorsive: il suo calcolo viene effettuato in una singola iterazione, considerando come relazioni EDB date tutte le occorrenze di predicati IDB che compaiono nella sezione *generate*, e con queste assunzioni le sue regole sono sempre al punto fisso.

---

[2]Una query posta sotto forma di clausola di Horn con quantificatori universali

Quando invece parliamo di 'valutazione della sezione *generate*' intendiamo una singola passata di $E^2$. In ogni momento le regole di questa sezione non sono necessariamente al punto fisso, e una valutazione corrisponde ad una singola applicazione dell'operatore di punto fisso, che aggiorna le relazioni correnti relative a predicati IDB, specializzandone i contenuti. Ciò realizza il meccanismo di passaggio da una soluzione parziale più generale ad una più in profondità nell'albero di backtracking.

In sintesi quest'algoritmo chiede a $E^2$ di valutare il programma, secondo i criteri che fisseremo nel seguito, e questo in risposta suggerisce se in quel momento la valutazione può proseguire (attraverso la valutazione della sezione *check*), o se è il caso di spingere $E^2$ su un'altra strada (**EE.ActualIterator++**). Una soluzione è completa se, una volta che la sezione *generate* raggiunge il punto fisso -cioè non possono essere derivati nuovi elementi che consentano di scendere in maggiore profondità nell'albero di ricerca- la variabile $fail$ nè la $fail^*$ vengono derivate dalla valutazione della sezione *check*, e inoltre tutte le complementazioni ottenute per 'guessing' risultano effettive.

## 3.3 Struttura dettagliata di Semi-naïf$^+$

Passeremo ora a descrivere in profondità ogni singola operazione che $E^2$ compie durante il processo di valutazione. Focalizzeremo l'attenzione soprattutto sulle caratteristiche a cui deve uniformarsi una possibile implementazione: Vedremo in dettaglio:

1. Che cosa succede quando si chiede a $E^2$ di valutare la sezione *generate*.

2. Come sono fatti i cosiddetti *Oggetti di iterazione* e in base a che criterio vengono allocati, incrementati e rimossi.

3. Come viene valutata la sezione *check*.

4. Come il ciclo principale influenza indirettamente le strutture dati private di $E^2$.

### 3.3.1 Gli oggetti di iterazione

Prima di passare a descrivere i compiti che $E^2$ è chiamato a svolgere, è necessario presentare la struttura dati più importante che questo ha a disposizione. Gli *oggetti di iterazione* svolgono infatti un ruolo centrale nel funzionamento di $E^2$. Un oggetto di iterazione viene costruito facendo riferimento a una relazione data, e svolge il compito operativo di costruire in base a questa una sequenza di relazioni di uscita, le quali rappresentano le possibili scelte che un singolo oggetto di iterazione può fare.

Un oggetto di iterazione non è direttamente fruibile dallo sviluppatore: gli oggetti gerarchicamente superiori possono inizializzarlo, scorrere lungo la sequenza di relazioni generata, esaminarne l'elemento corrente, verificare se la sequenza è terminata.

Volendo definire un oggetto di iterazione, facendo uso del concetto di classe proprio della programmazione OOP, allora diremo che

**Definizione 3.1** *( la classe Iterator ) Un oggetto di iterazione (o più semplicemente* iteratore*) è una classe che implementi i seguenti metodi:*

$$\begin{array}{lll} virtual & void & \textbf{init}(table\,\mathcal{R}) \\ virtual & bool & operator++ \\ virtual & table & \textbf{current}() \end{array}$$

Dove per *table* si intende l'istanza di una relazione di qualsiasi arità Pur essendo la definizione del tutto generale, è chiaro che non tutte le classi derivate conformi a quest'interfaccia saranno corrette. Noi ci limiteremo a considerare alcuni iteratori predefiniti, di cui esploreremo in dettaglio sia gli aspetti pratici che teorici, ma lasciando ampio spazio a future aggiunte di altre classi derivate da **Iterator**. Chiamiamo $\mathcal{U}$ l'universo di Herbrand delle costanti associate al programma $\mathsf{SKY}^{Plain}$ e alle relazioni di ingresso, e sia $\mathcal{B}$ la relativa base di Herbrand; tutti gli iteratori che ci limiteremo a considerare possiedono i seguenti requisiti:

- inizializzati su una data relazione $\mathcal{R}$ il metodo **current()** ritorna sempre una relazione $O_\mathcal{R}$ (che chiameremo *relazione corrente* o *relazione attuale* dell'iteratore) che è in qualche modo funzionale a $\mathcal{R}$ e di cardinalità complessiva polinomiale rispetto alla dimensione di $\mathcal{U}$.

- l'operazione '++' sposta il valore di **current()** su una nuova relazione $O'_\mathcal{R}$ diversa da tutte quelle precedentemente prodotte dallo stesso. Ritorna *true* se questa operazione può essere ripetuta con successo, *false* altrimenti. Non è lecito utilizzare '++' dopo la sua prima risposta negativa.

- '++' ritorna *true* al massimo un numero esponenziale di volte rispetto alla cardinalità di $\mathcal{U}$, cioè il numero massimo di possibili $O_\mathcal{R}$ che il metodo **current** può produrre è esponenziale, ma soprattutto finito.

In particolare, se chiamiamo $\mathcal{R}$ la relazione con cui viene inizializzato un dato iteratore, gli iteratori che considereremo sono[3]:

1. **range**, che ritorna come relazione attuale un'unica tupla di $\mathcal{R}$, prendendole in esame tutte.

2. **partition**, le cui relazioni attuali spaziano di volta in volta su tutte le possibili partizioni di cardinalità data di $\mathcal{R}$.

3. **permutation**, dove **current()** istanzia di volta in volta una relazione creata permutando gli elementi di $\mathcal{R}$, etichettando ciascuna tupla con un numero d'ordine.

---
[3]Si faccia attenzione a non confondere concettualmente questi nomi con quelli analoghi associati a un *costruttore di iterazione*. Vedremo che un costruttore di iterazione fa riferimento alla *potenziale* istanziazione di uno o più iteratori.

4. **subset**, che si occupa di scegliere uno tra tutti i possibili sottoinsiemi creati con le tuple di $\mathcal{R}$.

5. **any**, che istanzia una e una sola tupla di $\mathcal{R}$, senza garantire nulla su quale sia la tupla prescelta; a differenza di **range**, un iteratore **any** non può essere incrementato, ma semplicemente assume un'unico valore (una tupla singola). Può essere utile in tutti quei casi in cui il processo risolutivo deve iniziare da una qualche tupla del dominio (ad esempio un qualsiasi nodo di un grafo), ma non importa quale (basta che sia una). Corrisponde a una scelta che non può essere rivista.

   E' prevista in futuro l'aggiunta di costruttori analoghi per gli altri iteratori ( *AnyPartition, AnyPermutation*, ecc. ), di significato analogo.

6. **something**, che differisce in parte dai comuni iteratori, poichè restituisce come relazione attuale tutte le possibili relazioni di arità pari a quella di $\mathcal{R}$, costruibili utilizzando l'intero universo $\mathcal{U}$.

### 3.3.2 I gestori di iterazione

Un gestore di iterazione si occupa di raccogliere una collezione di oggetti di iterazione afferenti allo stesso costruttore. E' principalmente costituito da un insieme di coppie, gestito con politica LIFO, $< signature, oggetto\_di\_iterazione >$, dove *signature* è una tupla di arità pari a quella delle variabili di split associate al costruttore di iterazione corrispondente, e da una relazione *table*.

Si tratta quindi di una struttura dati che incapsula uno o più oggetti di iterazioni al suo interno.

Ogni gestore di iterazione è associato ad un costruttore di iterazione presente nel programma. Se questo costruttore non possiede variabili di split, allora il gestore di iterazione fa riferimento a un singolo oggetto di iterazione. Se invece il costruttore presenta delle variabili di split, il gestore di iterazione si occupa di creare un oggetto di iterazione per ogni diverso valore delle variabili di split con cui viene interrogato: ogni oggetto di iterazione viene marcato con una *signature* che non è altro che il valore delle variabili di split a cui corrisponde l'iteratore.

La relazione *table* accorpa tutti gli oggetti di iterazioni incapsulati nel gestore in un unica tabella comune, che sintetizza tutte le scelte fatte per un certo costruttore di iterazione ad un dato momento della valutazione.

### 3.3.3 Allocazione degli iteratori

Il primo passo che compie $E^2$ è quello di decidere se e quali oggetti di iterazione debbano essere allocati aggiornando di conseguenza le proprie strutture dati. Si prendono dunque in considerazione tutti i costruttori di iterazione partendo da quello più a sinistra; sia

$$iter_i(\mathbf{Y})[pred(\mathbf{X})](\alpha) \tag{4}$$

uno di questi, dove $\mathbf{Y} = Y_1 \ldots Y_m$, $\mathbf{X} = X_1 \ldots X_n$, $\alpha = \alpha_1 \ldots \alpha_k$.

Ad esso associamo un gestore di iterazione, con la relativa *table* inizialmente vuota. La semantica richiede di avere che l'intero vettore **Y** sia costituito da argomenti ritenuti *safe* a sinistra.

La definizione di *safeness* a sinistra coincide con quella classica esclusa una differenza fondamentale. Anzichè riferita a nomi di variabili, la sicurezza a sinistra riguarda singole occorrenze di variabili: ad esempio l'occorrenza di una variabile in un certo punto di un predicato può essere *safe* a sinistra, mentre un'altra no.

**Definizione 3.2** Safeness a sinistra. *Supponiamo di avere un certa occorrenza $\overline{X}$ di una variabile $X$ in un $pred_i$ di una regola come la 5. Diremo che $\overline{X}$ è sicura a sinistra se vale una delle seguenti affermazioni:*

1. *Esiste alla sinistra di $pred_i$ un predicato ordinario in cui compare la stessa variabile $X$.*

2. *Alla sinistra di $pred_i$ si trova un predicato predefinito $X = a$ dove $a$ è una costante, oppure vi si trova un predicato $X = Y$ e $Y$ è safe a sinistra.*

3. *Alla sinistra di $pred_i$ si trova un costruttore di iterazione $iter_i(\mathbf{Y})[pred(\mathbf{X})](\alpha)$ e $X$ occorre o in $\mathbf{X}$ o in $\alpha$.*

Ad esempio se avessimo la regola:

$$\begin{aligned} exit \leftarrow \ & \textbf{Permutation}(X)[scheduled(T,P)](N), \\ & task(X), \\ & \textbf{Range}(X)[processor(Y)] \end{aligned}$$

La $X$ che compare nel costruttore **Permutation** non è safe a sinistra, mentre la $X$ che occorre in **Range** lo è. E' evidente che se un certa occorrenza di variabile è safe a sinistra, lo saranno tutte le occorrenze della stessa variabile in predicati ancora più a sinistra.

Per avere un'ordine di allocazione degli iteratori facilmente controllabile dal programmatore, la semantica di SKY non accetta regole come quella appena vista. E' per questo che si è deciso di impedire che gli argomenti di split possano essere liberi: in questo modo il numero di oggetti di iterazione che ogni gestore può possedere è limitato dai valori che assumono i predicati ordinari nella stessa regola, ma più a sinistra.

Ad esempio è proibita la regola:

$$node(N) \leftarrow Range(N)[edge(X,Y)]$$

poichè la tabella di riferimento per il costruttore di iterazione dovrebbe avere un'entrata per ogni costante dell'universo di Herbrand.

Inoltre, poichè l'ordine con cui vengono fatte le scelte non deterministiche è da sinistra a destra, si è introdotto il concetto di sicurezza a sinistra. Queste due regole sono semanticamente diverse, ma entrambe valide:

$$e \leftarrow h(X), Subset(X)[p(K)], g(X)$$
$$e \leftarrow g(X), Subset(X)[p(K)], h(X)$$

Se supponiamo che al momento della valutazione le relazioni rispettivamente associate ad $h$ e a $g$ siano

| h | val |
|---|-----|
|   | a   |
|   | b   |

| g | val     |
|---|---------|
|   | alfa    |
|   | bravo   |
|   | charlie |

le tabelle dei gestori di iterazione ottenute dalla valutazione delle due regole differiscono, poichè una dipende dai valori di $h$, e l'altra dai valori di $g$:

| subseth | # | val   |
|---------|---|-------|
|         | a | Zeta  |
|         | a | Zappa |
|         | b | Zeta  |
|         | b | Zappa |

| subsetg | #       | val   |
|---------|---------|-------|
|         | alpha   | Zeta  |
|         | alpha   | Zappa |
|         | bravo   | Zeta  |
|         | bravo   | Zappa |
|         | charlie | Zeta  |
|         | charlie | Zappa |

Supponendo quindi che ogni costruttore di iterazione usato possegga questi requisiti, $E^2$ è in grado di valutare le regole di $\mathsf{SKY}^{Plain}$ come se fossero quelle di un normale programma DATALOG, se si eccettua che un atomo, che sia un costruttore di iterazione può essere sottoposto esclusivamente all'operazione, che andiamo a definire, di *join esteso*.

**Definizione 3.3** Join esteso ( operatore $\bowtie_+$ ). *Sia data una relazione $p_i(\mathbf{P})$ - dove $\mathbf{P}$ rappresenta una qualche denominazione dei campi di $p_i$, e un costruttore di iterazione $iter_i(\mathbf{Y})[pred(\mathbf{X})](\alpha)$; sia la sequenza di variabili $\mathbf{Y}$ contenuta in $\mathbf{P}$.*

*Definiremo*

$$out(\mathbf{Z}) = pred_i(\mathbf{P}) \bowtie_+ iter_i(\mathbf{Y})[pred(\mathbf{X})](\alpha)$$

*secondo i seguenti criteri: chiamiamo*

$$G(\mathbf{Y}) \equiv \prod_{\mathbf{Y}} pred_i(\mathbf{P})$$

*cioè sia $G$ la relazione ottenuta tramite la proiezione di $pred_i$ sui campi in cui compare la stessa variabile sia in $\mathbf{X}$ che in $\mathbf{Y}$. Sia BAG il gestore di iterazione associato a $iter_i$. Se $G \neq \emptyset$, per ogni tupla $g$ di $G$ si compiano le seguenti operazioni:*

1. *Se in* BAG *non esiste un oggetto di iterazione con* signature *pari a g allora lo si crei; sia questo del tipo* iter$_i$: *lo si inizializzi in base al valore della relazione* pred; *sia* iter *tale oggetto.*

2. *Si aggiorni BAG.table in questo modo:*

$$BAG.table = BAG.table \cup g \times iter.current()$$

*Se il costruttore di iterazione non possiede variabili di split, ci si comporti così:*

1. *Se in* BAG *non esiste un oggetto di iterazione allora lo si crei del tipo* iter$_i$ *e in base al valore della relazione* pred. *Sia* iter *tale oggetto.*

2. *Si aggiorni BAG.table in questo modo:*

$$BAG.table = iter.current()$$

*A questo punto diremo che*

$$out(\mathbf{Z}) = pred_i(\mathbf{P}) \bowtie BAG.table(\mathbf{Y}, \mathbf{X}, \alpha)$$

*Dove* $\mathbf{Z}$ *è costituito dall'unione senza ripetizioni di* $\mathbf{P}$, $\mathbf{Y}$, $\mathbf{X}$ *e* $\alpha$.

E' importante notare che, a differenza dell'operazione di join classica, il join esteso, anche e soprattutto per gli effetti collaterali che comporta sulle strutture dati che stanno alla base di un costruttore di iterazione, non è in generale nè commutativo nè associativo.

**3.3.3.1 Come avviene una valutazione della sezione *generate*** L'$E^2$ è il cuore di ciascuna iterazione di Semi-naïf$^+$: quando viene invocata una iterazione di valutazione, $E^2$ svolge il compito di acquisire il descrittore corrente di progresso, che in sostanza è costituito da un'istanza attuale dei predicati coinvolti nel programma (e rappresenta la soluzione parziale in esame), e di modificarlo nel tentativo di scendere in profondità nell'albero di backtracking. Le strutture dati a disposizione di $E^2$ sono:

1. Uno stack $\mathcal{S}$ di coppie $< gestore\_di\_iterazione, oggetto\_di\_iterazione >$.

2. Un database $\mathcal{D}$ che supporremo sia una raccolta di terne $T = < \mathcal{R}, t, \textit{ref} >$ dove $\mathcal{R}$ indica il nome della relazione cui appartiene la tupla $t$ e *ref* rappresenta una sequenza, eventualmente vuota, di riferimenti a degli oggetti di iterazione. Anche $\mathcal{D}$ potrà essere implementato con struttura a pila, a discrezione del progettista del sistema.

In sostanza $\mathcal{D}$ costituirà la rappresentazione estensionale attualmente raggiunta dai predicati in gioco nel programma SKY$^{Plain}$, arricchita dal riferimento agli oggetti di iterazione che hanno determinato la presenza di ciascuna tupla, e un cui cambiamento ne potrebbe inficiare la validità. Lo stack $\mathcal{S}$ è invece un raccolta di oggetti di iterazione, che potremmo considerare dei descrittori di 'spazi di guessing'.

Dato il programma $\mathsf{SKY}^{Plain}$, il passo preliminare da fare è quello di determinare una stratificazione delle regole rispetto al metapredicato $CO$; si può a questo scopo utilizzare un algoritmo come il 3.5 di [Ull91], tenendo conto che un costruttore di iterazione fa riferimento allo strato relativo al predicato argomento. Supporremo quindi di aver potuto determinare l'esistenza di una stratificazione tramite un algoritmo del genere, e di possedere così, per ciascun predicato del programma $\mathsf{SKY}^{Plain}$, un intero che indichi lo strato di appartenenza.

Se non è possibile determinare una stratificazione, il sistema definitivo potrà comunque proseguire l'elaborazione chiedendo all'utente di trasformare opportunamente alcuni tra i predicati $co$ rilevatisi critici in $co*$, o, secondo una strategia automatica di cui non ci occuperemo per il momento, provvedendo esso stesso alla modifica del programma.

Dunque, attribuiremo a ciascuna regola un primo ordine di valutazione, partendo dalle regole il cui predicato di testa appartiene allo strato più basso (valori di $i$ minori); a parità di strato, $E^2$ valuterà le regole nell'ordine dato dal grafo delle dipendenze, e infine nell'ordine in cui l'utente le ha volute specificare nel sorgente. Bisogna tenere presente che occorrerà raggiungere il punto fisso per le regole dello strato corrente prima di passare allo strato successivo.

Si tenga presente che, mentre questo criterio è meno rilevante per l'algoritmo Semi-naïf classico, riveste invece una certa importanza per quanto riguarda l'allocazione degli oggetti iterazione; quindi, per ottenere l'albero di backtracking voluto, sarà importante l'ordine di valutazione con cui si è specificato il problema.

Supponiamo allora di voler valutare una certa regola $\mathsf{SKY}^{Plain}$: questa possiederà, oltre a dei predicati ordinari o predefiniti, anche dei costruttori di iterazione:

$$head(\mathbf{X}) \leftarrow pred_1(\mathbf{X}_1) \ldots pred_n(\mathbf{X}_n) \qquad (5)$$

Dove $pred_1 \ldots pred_n$ possono essere sia predicati ordinari che costruttori di iterazione, mentre $head$ è un predicato ordinario. $E^2$ compie l'operazione di valutazione di una regola come la 5 in maniera formalmente identica alla procedura **EVAL-RULE-INCR** descritta in [Ull91], tenendo conto di alcune differenze chiave, che permettono di trattare i costruttori di iterazione, non significativi per l'algoritmo Semi-naïf classico:

1. Le operazioni di Join devono svolgersi da sinistra a destra.

2. Ogni operazione di Join che coinvolga un costruttore di iterazione deve usare il Join esteso anzichè il comune natural-join.

3. Se il primo sottoobiettivo di una regola è proprio un costruttore di iterazione, l'unico modo in cui possa essere sicuro a sinistra è che il vettore $\mathbf{Y}$ sia vuoto (Supponiamo che sia allora $I \equiv iter_i[pred(\mathbf{X})](\alpha)$). Per risolverlo, si usi al suo posto la relazione $\mathrm{ITER\_PRED}_k(\mathbf{X}, \alpha)$ ottenuta allocando un opportuno oggetto di iterazione *iter* nel gestore di iterazione di $I$, se questo non esiste già, e assegnando

$$ITER\_PRED_k = iter.current()$$

Inoltre la procedura $EVAL-INCR$ non considera, per quanto riguarda la valutazione incrementale, i costruttori di iterazione, cioè dovremo escludere dalla lista $S_1, \ldots, S_n$ non solo i predicati predefiniti, ma anche i costruttori di iterazione.

Supponiamo ad esempio di dover valutare la regola:

$$h(X,Y) \leftarrow r(X,Y), g(Y,Z), Subset(Z)[f(X,Y)]$$

L'espressione di algebra relazionale relativa al corpo di questa clausola sarà:

$$(R(X,Y) \bowtie G(Y,Z)) \bowtie_+ Subset(Z)[f(X,Y)]$$

dove le operazioni vanno eseguite rigorosamente nell'ordine dato dalle parentesi.

Chiamiamo $EVAL-INCR'$ una procedura così modificata. Quando essa viene invocata da $E^2$ per valutare una regola, si avrà come risultato diretto una relazione $\mathcal{R}$ che andrà aggiunta al database $\mathcal{D}$: ogni tupla di $\mathcal{R}$ sarà però marcata in maniera tale da indicare da quale regola dipende la sua presenza. Vediamo in dettaglio:

**Algoritmo 3.2** *Struttura di una singola passata di $E^2$. Sia $m$ il numero di regole relativo allo strato corrente facenti parte della sezione* generate, *e $k$ il numero di predicati intensionali o appartenenti allo strato precedente; supponiamo di avere assegnato a ciascuna regola un numero d'ordine secondo i criteri sopra visti. $G_1, \ldots, G_k$ saranno le relazioni attuali dei predicati intensionali presenti nel programma, che chiameremo $p_1, \ldots, p_k$; $\Delta G_1, \ldots, \Delta G_k$ saranno invece* relazioni incrementali *associate agli stessi; data una regola $g_i$ indichiamo inoltre con $g_i^h$ (risp. $\Delta g_i^h$) la relazione (una tra $G_1, \ldots, G_k$ risp. $\Delta G_1, \ldots, \Delta G_k$) associata al predicato di testa.*

PROC $E^2$ generic iteration;

    FOR j := 1 to k DO BEGIN
       $\Delta Q_j := \Delta G_j$;
       $\Delta G_j := \emptyset$;
    END;
    FOR i := 1 TO m DO BEGIN
       $\Delta G$ := EVAL-INCR$'(g_i, P_1, \ldots, P_k, G_1, \ldots, G_k, \Delta Q_i, \ldots, \Delta Q_k)$;
       $\Delta G := \Delta G - g_i^h$;
       $\Delta g_i^h := \Delta g_i^h \cup \Delta G$;
       Per ogni tupla $t$ di $\Delta G$ aggiungi a $\mathcal{D}$ la coppia $<t, ref>$,
       dove $ref$ e' un riferimento a tutti gli oggetti di iterazione
       allocati nella EVAL-INCR$'$ appena eseguita;

Aggiungi tutti gli oggetti di iterazione che sono stati creati nello stack
  $\mathcal{S}$ nell'ordine in cui sono stati allocati;
END;
FOR j := 1 to k DO
  $G_i := G_i \cup \Delta G_i$;
  FixedPointForThisLayer := $\Delta G_i = \emptyset \ \forall i$;
ReachedFixedPoint := FixedPointForThisLayer $\wedge$ (questo e' l'ultimo strato);
IF (FixedPointForThisLayer) THEN CurrentLayer++;

Nel caso in cui sia la prima volta che si affronta un certo strato, non ci si potrà avvalere delle relazioni incrementali per cui, nella prima iterazione, le cose andranno in maniera leggermente diversa:

PROC $E^2$ first iteration on a layer;

FOR i := 1 TO m DO BEGIN
  $\Delta G := \text{EVAL}(g_i, P_1, \ldots, P_k, \emptyset, \ldots, \emptyset)$;
  $\Delta g_i^h := \Delta g_i^h \cup \Delta G$;
  Per ogni tupla $t$ di $\Delta G$ aggiungi a $\mathcal{D}$ la coppia $<t, ref>$,
  dove $ref$ e' un riferimento a tutti gli oggetti di iterazione
  allocati nella EVAL$'$ appena eseguita;
  Aggiungi tutti gli oggetti di iterazione che sono stati creati nello stack
  $\mathcal{S}$ nell'ordine in cui sono stati allocati;
  $G_i := \Delta G_i$;
END;
FixedPointForThisLayer := $\Delta G_i = \emptyset \ \forall i$;
ReachedFixedPoint := FixedPointForThisLayer $\wedge$ (questo e' l'ultimo strato);
IF (FixedPointForThisLayer) THEN CurrentLayer++;

**3.3.3.2 Come $E^2$ incrementa e dealloca gli oggetti di iterazione** Ci occuperemo qui del meccanismo con cui $E^2$ aggiorna le sue strutture dati quando il 'guess' corrente fallisce per cui si innesca il meccanismo di backtracking.

Una chiamata di **ActualIterator++** comporta le seguenti operazioni (chiamiamo $top$ l'oggetto di iterazione in cima a $\mathcal{S}$):

- Si cerca di incrementare $top$.

- Si rimuovono tutte le tuple di $\mathcal{D}$ che presentano il riferimento a $top$. Se la rimozione interessa una qualche tupla che inerente ad un predicato di uno strato precedente a quello corrente, si decrementa $CurrentLayer$.

- Se non si è potuto incrementare $top$ (tutti i guess ad esso afferenti sono stati tentati) lo si rimuove da $\mathcal{S}$.

- Si restituisce $true$ o $false$ a seconda del successo nell'operazione di incremento di $top$.

**EE.Iterators.empty()** ritorna invece lo stato attuale ( presenza o meno di oggetti iterazione) di $\mathcal{S}$.

**3.3.3.3 Come viene valutata la sezione *check*** La sezione *check* è di fatto un semplice programma DATALOG con negazione stratificata, senza ricorsione. Può essere valutato in una sola passata tramite algoritmo Semi-naïf. Ogni predicato occorrente anche nella sezione *generate* viene considerato come EDB, prendendone come valore la relazione corrente presente in $\mathcal{D}$. Non è consentita la presenza in testa a una regola di un predicato occorrente anche nella sezione *generate*.

# 4 Sintesi di algoritmi con SKY

# 5 Il problema del circuito hamiltoniano

Questo storico problema si presta molto bene a strategie di calcolo tramite backtracking, ma ci consentirà di mostrare come SKY possa essere usato altrettanto bene per specificare algoritmi enumerativi.

**Esempio 5.1 (Hamiltonian Circuit)** *Dato un grafo diretto $G = <V, E>$, questo possiede un circuito Hamiltoniano? E' possibile cioè trovare un percorso che, partendo da un qualsiasi $V_0 \in V$, tocchi tutti i nodi una e una sola volta tornando al punto di partenza?*

Fissiamo il nostro dominio di valutazione con il predicato $ciclo(X,N)$, dove X identifica un nodo, e N il numero d'ordine che un nodo assume nella sequenza. Un ciclo sarà valido se soddisfa la seguente sezione *check* senza derivare fail:

$$
\begin{align}
fail &\leftarrow ciclo(X,K), ciclo(X,J), J \neq K \tag{6} \\
fail &\leftarrow ciclo(X,K), ciclo(X,K+1), CO[edge(X,Y)] \tag{7} \\
fail^* &\leftarrow ciclo(X,1), ciclo(Y,N), CO[edge(X,Y)] \tag{8} \\
usato(X) &\leftarrow ciclo(X,Some) \tag{9} \\
fail^* &\leftarrow node(X), CO[usato(X)] \tag{10} \\
&\tag{11}
\end{align}
$$

Limitiamo inoltre ciascun ciclo a

$$
\begin{align*}
&[bounds] \\
ciclo(X,N) &\leftarrow node(X), \{1..count<node>\}(N)
\end{align*}
$$

A questo punto possiamo scegliere se interrogare il sistema usando l'enumerazione, e per far questo basta la regola statica:

$$ciclo(X,N) \leftarrow permutation[node(X)](N)$$

C'è da osservare che tale regola di generazione rende inutile il vincolo 11. Possiamo invece pensare di creare il circuito in base a dei circuiti parziali, in cui le ramificazioni dell'albero di ricerca corrispondono ai possibili nodi che possono essere aggiunti ad una data iterazione nella ricerca di un circuito completo:

$$ciclo(X, 1) \leftarrow any[node(X)]$$
$$ciclo(X, N+1) \leftarrow ciclo(Y, N), range(N)[edge(Y, X)]$$

In questo caso c'è da osservare che bisogna mantenere il vincolo 8, poichè l'iteratore *range* sceglie uno tra tutti i possibili archi, per poi verificare eventualmente che questo non può legarsi a nessuna delle tuple di *ciclo*. Il processo di backtracking è comunque corretto, grazie al vincolo 11, e possiamo capirlo meglio con il seguente esempio:

**Esempio 5.2 (Casi particolari di backtracking)** *supponiamo di voler verificare che il seguente grafo è hamiltoniano:*

node

| Name |
|---|
| Aurora |
| Solaria |
| Terra |

edge

| From | To |
|---|---|
| Aurora | Solaria |
| Solaria | Terra |
| Terra | Aurora |

e di trovarci a un certo punto della valutazione con la situazione

ciclo

| Node | # |
|---|---|
| Aurora | 1 |
| Solaria | 2 |

rangeedge1

| # | From | To |
|---|---|---|
| 1 | Aurora | Solaria |

rivalutiamo il programma *generate* in base alla situazione corrente: l'iteratore *rangeedge1* viene per la prima volta invocato con un valore di $N = 2$ e dunque viene chiamato a fare una scelta tra gli archi. Viene selezionato l'arco (Aurora,Solaria). La situazione diventa:

ciclo

| Node | # |
|---|---|
| Aurora | 1 |
| Solaria | 2 |

rangeedge1

| # | From | To |
|---|---|---|
| 1 | Aurora | Solaria |
| 2 | Aurora | Solaria |

Come si vede, non viene aggiunto nessun nodo al circuito parziale, poiche l'arco selezionato non parte da Solaria. Il programma di *check* valida questa situazione come plausibile. Ma all'iterazione successiva sul programma *generate* ci si rende conto che nessuna regola porta alla generazione di nuovi nodi, poichè il programma *generate* raggiunge il punto fisso. Quando il sistema raggiunge il punto fisso, assume di trovarsi su un potenziale circuito completo, per cui cerca di validarlo in base al valore del letterale $fail^*$: in questo caso la situazione corrente non è evidentemente un ciclo completo, e infatti viene derivato $fail^*$ con la regola 11. Il meccanismo innesca a questo punto il processo di backtracking: in pratica vengono annullate le iterazioni precedenti fino al punto in cui si è fatta l'ultima scelta non deterministica, cioè si torna alla situazione iniziale del nostro esempio. La sezione *generate* viene rivalutata, e a questo punto l'iteratore *Range* prova con un altro guess:

ciclo

| Node | # |
|---|---|
| Aurora | 1 |
| Solaria | 2 |
| Terra | 3 |

rangeedge1

| # | From | To |
|---|---|---|
| 1 | Aurora | Solaria |
| 2 | Solaria | Terra |

La sezione *check* convalida questa situazione e prova di nuovo a incrementare il ciclo con un'altra valutazione di *generate*. Ma i limiti aritmetici bloccano il calcolo ( non si può andare oltre il terzo nodo), per cui viene individuato un punto fisso: questo è riconosciuto come un circuito hamiltoniano, poichè possiede tutti i nodi, gli estremi sono congiunti, e nessun nodo è usato due volte. L'algoritmo SKY risponde 'YES' per questa istanza. Riportiamo infine il programma completo:

$$
\begin{aligned}
&[bounds] \\
ciclo(X, N) &\leftarrow node(X), \{1..count<node>\}(N) \\
&[generate] \\
ciclo(X, 1) &\leftarrow any[node(X)] \\
ciclo(X, N+1) &\leftarrow ciclo(Y, N), range(N)[edge(Y, X)] \\
&[check] \\
fail &\leftarrow ciclo(X, K), ciclo(X, J), J \neq K \\
fail &\leftarrow ciclo(X, K), ciclo(X, K+1), CO[edge(X, Y)] \\
fail^* &\leftarrow ciclo(X, 1), ciclo(Y, N), CO[edge(X, Y)] \\
usato(X) &\leftarrow ciclo(X, Some) \\
fail^* &\leftarrow node(X), CO[usato(X)]
\end{aligned}
$$

## 5.1 I metaalgoritmi di enumerazione, backtracking e SKY

Il programma SKY appena visto ci consente di fare alcune considerazioni circa la corrispondenza tra una specifica SKY e i metaalgoritmi di enumerazione e/o backtracking tradizionali. Possiamo ad esempio analizzare il metaalgoritmo di enumerazione:

```
void Enumerazione(istanza i)

{ for (i.Primo(); !i.Ultimo(); i.successore())
  if (i.ammissibile())
    { i.OutputYES();
    return;
    }
  i.OutputNO();
}
```

Questo metaalgoritmo può essere istanziato da un programma SKY che possieda una sezione *generate* non ricorsiva, mentre la sezione *check* non deve fare uso di $fail$, ma solo di $fail^*$. Il principale modo di fare backtracking in SKY -la valutazione incrementale- è così disabilitato, come anche i meccanismi di filtraggio sui nodi intermedio: la sezione *check* si limita a convalidare solo soluzioni finali.

Un programma completo per il ciclo hamiltoniano che faccia uso della sola enumerazione è:

[generate]

$$\begin{aligned}
ciclo(X,N) &\leftarrow permutation[node(X)](N) \\
[check] & \\
fail^* &\leftarrow ciclo(X,K), ciclo(X,J), J \neq K \\
fail^* &\leftarrow ciclo(X,K), ciclo(X,K+1), CO[edge(X,Y)] \\
fail^* &\leftarrow ciclo(X,1), ciclo(Y,N), CO[edge(X,Y)]
\end{aligned}$$

Tutte le entità a cui fa riferimento la tecnica enumerativa sono facilmente evidenziabili: lo spazio di ricerca coincide con l'insieme di tutti i potenziali cicli che possono essere valutati. La valutazione della sezione *generate* è sempre al punto fisso, per cui ci si sposta *esaustivamente* sullo spazio di ricerca:

- **primo()** può essere visto come la prima valutazione della sezione *generate* che predispone gli iteratori allocandoli staticamente, in questo caso si tratta di esplorare tutte le permutazioni possibili dei nodi;

- **ultimo()** corrisponde all'esaurimento di tutti gli iteratori allocati (allocazione che è avvenuta unicamente alla prima iterazione).

- **successore()** corrisponde ad un'incremento dell'iteratore $Permutation$ innescato dal fatto che, trovandosi il motore di valutazione al punto fisso, la derivazione di $fail^*$ nel valutare la sezione *check* impone di aggiornare gli iteratori per esplorare altri elementi dello spazio di ricerca.

- **ammissibile()** corrisponde evidentemente a una valutazione della sezione *check* dove si risponde *true* se $fail^*$ viene derivato.

Se prendiamo invece in considerazione il tradizionale metodo di backtracking, ci sono alcune piccole differenze che rendono SKY più generale; in particolare con SKY è molto evidente la distinzione tra criteri di pruning *necessari* -cioè relativi alla funzione di ammissibilità dei nodi intermedi dell'albero di ricerca- e vincoli di ammissibilità dei soli nodi foglia (soluzioni complete).

Come già sappiamo se si vuole attivare il meccanismo di backtracking, in SKY bisogna fare uso della ricorsione -che viene intesa come un'operazione di *prossimo passo*- e del letterale $fail$ non asteriscato. Detto questo si può subito osservare che rispetto al metodo di backtracking tradizionale, SKY istanzia i metodi ivi usati in questo modo:

1. **Violavincoli()** è chiaramente la valutazione della sezione *check*, e risponde *true* se viene derivato $fail$.

2. **Scendilivello()** è rappresentato da una valutazione della sezione *generate*.

3. l'utente è sollevato in SKY dall'esprimere l'analogo di **Salilivello()**, direttamente: il sistema riconosce autonomamente quando l'iteratore che fa riferimento al nodo corrente è esaurito, e rimuove tutte le strutture dati ad esso collegate.

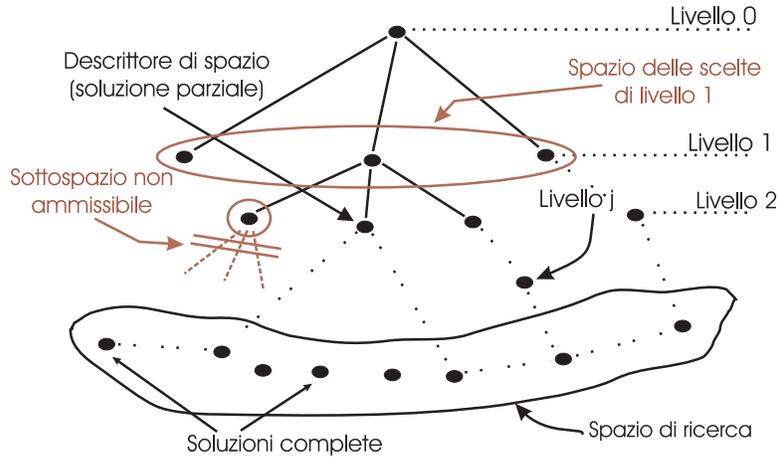

Figura 1: Struttura di un tipico albero di ricerca generato da $E^2$, riferito in particolare all'esempio 5.1.

4. **SoluzioneCompleta()** corrisponde al riconoscimento del punto fisso della sezione *generate*.

5. Anche i meccanismi di **PrimoStessoLivello()** e **ProssimoStessoLivello**, fanno riferimento all'inizializzazione e incremento dell'iteratore corrente, procedimenti del tutto trasparenti all'utente in SKY.

In figura 5.1 mostriamo come è strutturato l'albero di ricerca per l'esempio 5.1.

C'è da evidenziare che SKY consente di sintetizzare anche algoritmi con tecniche di valutazione mista, come vedremo nella prossima sezione.

# 6 Il tempo minimo di diffusione di un messaggio in broadcast

**Esempio 6.1 (Minimum broadcast time [GJ79])** *Supponiamo di avere un grafo $G = <V, E>$, un sottoinsieme dei suoi nodi $V_0 \subseteq V$ e un intero positivo K. Ci chiediamo: può un messaggio essere 'diffuso' dall'insieme di partenza $V_0$ a tutti i vertici in tempo K (considerando il tempo di propagazione attraverso un arco come unitario), cioè, c'è una sequenza di insiemi $V_0, E_1, V_1, E_2, \ldots, E_K, V_K$ tale che ogni $V_i \subseteq V$, ogni $E_i \subseteq E$, $V_K = V$, e, per $1 \leq i \leq K$, a) ogni arco in $E_i$ ha esclusivamente un estremo su un nodo di $V_{i-1}$, b) nessun arco di $E_i$ condivide un estremo, e c) $V_i = V_{i-1} \cup \{v : (u,v) \in E_i\}$?*

Useremo il predicato $LayerEdge(Num, From, To)$ per indicare l'appartenenza di un arco a un certo strato nel processo di broadcast, e $LayerNode(Num, Name)$ per indicare l'appartenenza di un nodo $Name$ allo strato $Num$: come al solito useremo i predicati $edge(X,Y)$ e $node(X)$ per rappresentare il nostro grafo di input, mentre $LayerNode_0(From, To)$ sarà il predicato EDB che contiene il sottoinsieme sorgente $V_0$. Stabiliamo innanzitutto i bounds su cui ciascun predicato si deve muovere:

$$[bounds]$$
$$LayerEdge(N, F, T) \leftarrow \{1..k\}(N), edge(F, T)$$
$$LayerNode(N, X) \leftarrow \{1..k\}(N), node(X)$$

Aggiungiamo subito come regole di check:

$$fail \leftarrow LayerEdge(N, F, T), LayerNode(N-1, F), LayerNode(N-1, T)$$
$$fail \leftarrow LayerEdge(N, F, T), LayerEdge(N, F, X), T \neq X$$

mentre possiamo usare il vincolo $c)$ per generare $LayerNode$ in base a $LayerEdge$:

$$LayerEdge(N, X, Y) \leftarrow LayerEdge(N, Y, X)$$
$$LayerNode(N+1, X) \leftarrow LayerNode(N, X)$$
$$LayerNode(N+1, X) \leftarrow LayerNode(N, X), LayerEdge(N+1, X, Y)$$

dove la prima regola ci serve per gestire con maggior sintesi il fatto che ci occupiamo di un grafo non diretto. La costruzione dinamica del nostro dominio è affidata alle regole:

$$LayerEdge(N, X, Y) \leftarrow Partition[edge(X, Y), k](N)$$

Ci resta infine da definire la condizione per cui una soluzione finale è ammissibile:

$$FinalNodes(X) \leftarrow LayerNode(k, X)$$
$$fail^* \leftarrow CO[FinalNodes(X)]$$

Il programma complessivo è il seguente:

$$[bounds]$$
$$LayerEdge(N, F, T) \leftarrow \{1..k\}(N), edge(F, T)$$
$$LayerNode(N, X) \leftarrow \{1..k\}(N), node(X)$$
$$[generate]$$
$$LayerEdge(N, X, Y) \leftarrow Partition[edge(X, Y), k](N)$$

$$
\begin{aligned}
LayerEdge(N, X, Y) &\leftarrow LayerEdge(N, Y, X) \\
LayerNode(0, X) &\leftarrow LayerNode_0(X) \\
LayerNode(N+1, X) &\leftarrow LayerNode(N, X) \\
LayerNode(N+1, X) &\leftarrow LayerNode(N, X), LayerEdge(N+1, X, Y) \\
[check]& \\
fail &\leftarrow LayerEdge(N, F, T), \\
&\quad LayerNode(N-1, F), \\
&\quad LayerNode(N-1, T) \\
fail &\leftarrow LayerEdge(N, F, T), \\
&\quad LayerEdge(N, F, X), \\
&\quad T \neq X \\
FinalNodes(X) &\leftarrow LayerNode(k, X) \\
fail^* &\leftarrow node(X), CO[FinalNodes(X)]
\end{aligned}
$$

E' interessante dunque, osservare che, in questo esempio, si fa un uso misto di backtracking ed enumerazione: la sezione di generazione fa un *guess* iniziale su tutta la partizione degli archi di $E$, mentre gli insiemi $V_i$ sono generati con regole ricorsive che innescano il backtracking ogniqualvolta non vengono rispettati i criteri che definiscono la diffusione *broadcast*. Il programma complessivo è sufficientemente efficiente.

## 7 Overloading inverso

L'uso dei template in SKY è molto vicino al medesimo concetto presente in C++. In poche parole si può pensare a un template come a un modo per definire un certo predicato IDB e il sottoprogramma che lo definisce, *a)* immediatamente nel posto in cui deve essere usato e *b)* in base a un *modello* generico riutilizzabile.

Ciò facilità la riutilizzabilità delle specifiche, ne migliora la leggibilità e la sinteticità. Un uso banale che ne potremmo fare è per esempio:

$$
\begin{aligned}
fail &\leftarrow collide<red> \\
fail &\leftarrow collide<blue> \\
fail &\leftarrow collide<green>
\end{aligned}
$$

Entrambe queste tre regole fanno riferimento allo stesso template

$template collide < color(1) > ()$

$$
collide \leftarrow edge(X, Y), color(X), color(Y)
$$

Quando si usa un template come quello visto, bisogna tenere presente che in realtà è come se si stesse scrivendo per esteso questo programma:

$$\begin{aligned} collide_{blue} &\leftarrow edge(X,Y), blue(X), blue(Y) \\ collide_{red} &\leftarrow edge(X,Y), red(X), red(Y) \\ collide_{green} &\leftarrow edge(X,Y), green(X), green(Y) \\ fail &\leftarrow collide_{blue} \\ fail &\leftarrow collide_{red} \\ fail &\leftarrow collide_{green} \end{aligned}$$

Abbiamo già visto che i possibili modi di unificare un template con una sua istanza, sono molteplici: in particolare, con un meccanismo che potremmo chiamare *overloading inverso*, si può invocare lo stesso template con predicati di arità diversa da quella specificata nel prototipo:

$$fail \leftarrow collide < coloring(\_, *, COL) >$$

unendo questa regola al template relativo, possiamo considerare il programma complessivo come una riscrittura di queste regole:

$$\begin{aligned} pcoloring(A,B) &\leftarrow coloring(A,\_,B) \\ collide_{coloring}(COL) &\leftarrow edge(X,Y), pcoloring(X,COL), pcoloring(Y,COL) \\ fail &\leftarrow collide_{coloring}(COL) \end{aligned}$$

Abbiamo deciso di chiamare questa modalità d'uso dei template *overloading inverso* perchè il meccanismo ricorda quello dell'overloading delle funzioni tipico del C++. La -enorme- differenza consiste nel fatto che, mentre in C++ l'overloading è un meccanismo per utilizzare lo stesso nome di funzione facendo capo a funzioni in realtà diverse, nel nostro caso si possono utilizzare forme di invocazioni diverse di un costruttore di template, per fare riferimento allo stesso unico prototipo.

Ovviamente i template sono di grande aiuto non solo per l'utente finale ma anche per fornire funzioni generiche di uso comune. Si può pensare che ad esempio il template di libreria *max* sia equivalente a:

$template max < p(\_) > (\_)$

$$\begin{aligned} exceeded(X) &\leftarrow p(X), p(Y), Y > X \\ max(X) &\leftarrow p(X), CO^*[exceeded(X)] \end{aligned}$$

Questo semplicissimo template è di uso completamente generale (senza problemi di stratificazione), e può essere usato per trovare il massimo di un argomento qualunque di un qualsiasi predicato (ad es. $max < pred(*, \_, *, *) > (X)$), raggruppare insiemi di valori massimi distinti (ad es. $max < pred(Y, Z, \_, *)(X)$), con l'uso di un'unico costruttore di template.

# 8 Altri esempi di specifica

### 8.0.1 Le k-regine

L'esistenza di un algoritmo risolutivo molto efficace non sminuisce gli spunti didattici che sorgono affrontando questo problema con la tecnica del backtracking.

**Esempio 8.1 (Le K-regine)** *Data una scacchiera di dimensione $k \times k$, si possono posizionare su questa $k$ regine, in maniera tale che nessuna di queste minacci un'altra regina, cioè si trovi o sulla stessa riga, o sulla stessa colonna, o sulla stessa diagonale?*

Useremo il predicato *pos* per indicare la posizione delle regine sulla scacchiera: il fatto $pos(a, b)$ indica che una regina è posizionata sulla colonna $a$, nella riga $b$. La strategia classica che specifichiamo è quella di aggiungere una regina per ogni colonna, eliminando quindi il vincolo di collisione su queste. Una soluzione parziale è rappresentata da un valore non definitivo della relazione associata a *pos*, che rappresenta una scacchiera non completa, ma con un certo numero di regine già posizionate e non in collisione.

$$
\begin{array}{rcl}
[bounds] & & \\
pos(X, Y) & \leftarrow & \{1..k\}(X), \{1..k\}(Y) \\
[generate] & & \\
pos(1, Y) & \leftarrow & range(X)[1..k(Y*)] \\
pos(X+1, Y) & \leftarrow & pos(X, SOME), range(X)[1..k(Y*)] \\
[check] & & \\
fail & \leftarrow & pos(X, Y), pos(X, Z), Y \neq Z \\
fail & \leftarrow & pos(X, Y), pos(Z, Y), X \neq Z \\
fail & \leftarrow & pos(X, Y), pos(J, K), J + K = X + Y, X \neq J, Y \neq K \\
fail & \leftarrow & pos(X, Y), pos(J, K), J + Y = X + K, X \neq J, Y \neq K \\
fail^* & \leftarrow & CO[pos(k, \_)]
\end{array}
$$

La sezione *check* specifica in sequenza i quattro vincoli che possono far fallire una soluzione parziale: collisione su una colonna, collisione su una riga, collisione sulla diagonale principale, collisione sulla diagonale secondaria. Da notare che il primo vincolo è superfluo nel nostro programma, perchè strutturalmente soddisfatto dalla sezione *generate*. Lo stesso, sarebbe invece necessario se avessimo adottato una specifica del dominio come:

$$pos(X,Y) \leftarrow Permutation[\{1..k\}(X)](Y)$$

### 8.0.2 Il set-splitting

**Esempio 8.2 (set splitting [GJ79])** *Sia data una collezione $C$ di sottoinsiemi di cardinalità 3 di un certo insieme $S$. Esiste una partizione di $S$ in due sottoinsiemi $S_1$ e $S_2$ tale che nessun sottoinsieme in $C$ è interamente contenuto o in $S_1$ o in $S_2$?*

Assumeremo che il predicato di input sia la relazione $c(X,Y,Z)$, che elenca tutti gli insiemi di 3 elementi appartenenti a $C$, mentre $S$ è codificato tramite il predicato $s(X)$. Scriveremo:

$$\begin{aligned} &[generate]\\ split(X,C) &\leftarrow partition[s(X),2](C)\\ &[check]\\ fail &\leftarrow c(X,Y,Z), split(X,C),\\ & \quad split(Y,C), split(Z,C) \end{aligned}$$

In questo esempio possiamo mostrare anche delle modalità di validazione dell'input: in particolare non tutte le relazioni di input $c(X,Y,Z)$ sono valide se i singoli elementi non appartengono ad $S$. Ciò può essere indicato con le regole:

$$\begin{aligned} fail &\leftarrow c(X,Y,Z), CO[s(X)]\\ fail &\leftarrow c(X,Y,Z), CO[s(Y)]\\ fail &\leftarrow c(X,Y,Z), CO[s(Z)] \end{aligned}$$

Inoltre $c$ indicherà dei sottoinsiemi di cardinalità 3 se e solo se ogni tupla è composta da elementi distinti:

$$\begin{aligned} fail &\leftarrow c(X,X,Z)\\ fail &\leftarrow c(X,Y,Y)\\ fail &\leftarrow c(X,Y,X) \end{aligned}$$

Questa ridondanza nella validazione dell'input è in parte dovuta alle carenze del modello relazionale, il cui tipo di dato principale è appunto la relazione.

C'è inoltre da osservare che è del tutto ridondante verificare questi vincoli se non al primo passo dell'algoritmo Semi-naïf$^+$. In generale, situazioni del genere possono essere ottimizzate introducendo una forma di valutazione differenziale della sezione *check*.

### 8.0.3 Rilevamento di errori in un circuito logico

Affronteremo infine un problema di una certa rilevanza pratica, onde mostrare la bontà di un sistema basato su SKY, sul terreno dell'utilizzo reale.

**Esempio 8.3 (Fault detection in logic circuits [GJ79])** *Sia dato un grafo diretto e aciclico $G = \langle V, A \rangle$, con un singolo vertice $v^* \in V$ senza archi uscenti e un solo arco entrante, un assegnamento $f : (V - \{v^*\}) \to \{I, and, or, not\}$ tale che $f(v) = I$ implichi che $v$ non ha archi entranti, $f(v) = not$ implichi che $v$ ha un solo arco entrante, e $f(v) = and$ o $f(v) = or$ implichi che $v$ ha due archi entranti, e un sottoinsieme $V' \subseteq V$.*

*Possono essere rilevati tutti gli errori occorrenti ad un singolo vertice di $V'$ semplicemente tramite prove di input-output, cioè: vedendo $G$ come un circuito con vertici di input $I$, vertice di output $v^*$, e con porte logiche per le funzioni 'and', 'or', e 'not' ai vertici specificati, c'è per ogni $v \in V'$ e $x \in \{T, F\}$ un assegnamento di un valore a ogni vertice in $I$ estratto da $\{T, F\}$ tale che l'output del circuito per tali valori di input differisce dall'output dello stesso circuito con l'output della porta al nodo $v$ 'bloccato' al valore $x$?*

Codificheremo ogni nodo con il predicato $gate(X, Type)$ dove $Type$ può essere uno tra $\{in, not, and, or, out\}$, mentre una porta $X$ connetterà il suo output in input alla porta $Y$ se vale $wire(X, Y)$. L'insieme di porte logiche di cui si vuole verificare la rilevabilità del guasto sarà codificato dal predicato $test(X)$. La relazione $out(X)$ certifica se in uscita ad un dato nodo si ha $T$. Imposteremo il problema come la ricerca di una porta logica il cui guasto sia riconoscibile. Non ci occupiamo per il momento del problema di validazione dell'input.

Con questo template codifichiamo l'interrogazione di un dato circuito. Si noti che il programma è localmente stratificato quando si ha in input un circuito valido, ma non stratificato, per cui siamo costretti comunque a fare uso della negazione $CO*$.

$template\ out < gate(2), wire(2), input(1) > ()$

$$
\begin{aligned}
out &\leftarrow wire(X, Y), gate(Y, out), out(X) \\
out(X) &\leftarrow gate(X, and), wire(Y_1, X), wire(Y_2, X), out(Y_1), out(Y_2), Y_1 \neq Y_2 \\
out(X) &\leftarrow gate(X, or), wire(Y_1, X), out(Y_1) \\
out(X) &\leftarrow gate(X, not), wire(Y_1, X), CO^*[out(Y_1)] \\
out(X) &\leftarrow gate(X, in), input(X)
\end{aligned}
$$

Ora dobbiamo codificare lo spazio dei circuiti di prova: un dato input è inutilizzabile per rilevare un guasto su un dato nodo se l'output è uguale con o senza il nodo attualmente preso in considerazione bloccato. Per cui affiancheremo di volta in volta al circuito dato come istanza, un circuito costruito eliminando la porta da esaminare ad una data iterazione.

$$
\begin{aligned}
&[generate] \\
tested(X) &\leftarrow range[test(X)] \\
input(X) &\leftarrow gate(X, in), subset[gate(X, T)] \\
gate'(X, T) &\leftarrow gate(X, T), CO[tested(X)] \\
gate'(X, in) &\leftarrow tested(X) \\
wire'(X, Y) &\leftarrow wire'(X, Y), CO[tested(X)] \\
input'(X) &\leftarrow input(X) \\
input'(X) &\leftarrow tested(X), Something \\
out_1 &\leftarrow out < gate(\_,\_), wire(\_,\_), input(\_) > \\
out_2 &\leftarrow out < gate'(\_,\_), wire'(\_,\_), input'(\_) > \\
&[check] \\
fail^* &\leftarrow out_1, out_2 \\
fail^* &\leftarrow CO[out_1], CO[out_2]
\end{aligned}
$$

C'è qui da sottolineare l'uso dell'iteratore *Something*, che se invocato come predicato di arità nulla, può spaziare solo tra i valori *true* e *false*: in questo modo ciascun circuito difettoso verrà testato bloccando la porta difettosa al valore 0 o al valore 1.

## Riferimenti bibliografici